\documentclass[aps,twocolumn,preprintnumbers,amsmath,superscriptaddress,amssymb,floats,nofootinbib]{revtex4-1}
\usepackage{graphicx,color,footnote}

\begin{document}

\def\Journal#1#2#3#4{{#1} {\bf #2}, #3 (#4)}
\def\NCA{\rm Nuovo Cimento}
\def\NPA{{\rm Nucl. Phys.} A}
\def\NIM{\rm Nucl. Instrum. Methods}
\def\NIMA{{\rm Nucl. Instrum. Methods} A}
\def\NPB{{\rm Nucl. Phys.} B}
\def\PLB{{\rm Phys. Lett.}  B}
\def\PRL{\rm Phys. Rev. Lett.}
\def\PRD{{\rm Phys. Rev.} D}
\def\PRC{{\rm Phys. Rev.} C}
\def\ZPC{{\rm Z. Phys.} C}
\def\JPG{{\rm J. Phys.} G}



\title{Reactor Antineutrino Anomaly with known $\theta_{13}$}
\date{\today}
\author{C. Zhang}\email[]{chao@bnl.gov}
\affiliation{Brookhaven National Laboratory, Upton, NY 11973, USA}
\author{X. Qian}\email[]{xqian@caltech.edu}
\affiliation{Kellogg Radiation Laboratory, California Institute of Technology, Pasadena, CA 91125, USA}
\author{P. Vogel}\email[]{pxv@caltech.edu}
\affiliation{Kellogg Radiation Laboratory, California Institute of Technology, Pasadena, CA 91125, USA}
\begin{abstract}
We revisit the reactor antineutrino anomaly using the recent reactor flux independent
determination of sizable $\theta_{13}$ by considering the full set of the absolute 
reactor $\bar{\nu}_e$ flux measurements. When normalized to the predicted flux of Mueller 
{\it et al.}~\cite{nspec2}, the new world average, after including results from Palo Verde, Chooz, 
and Double Chooz, is 0.959 $\pm$ 0.009 (experiment uncertainty) $\pm$ 0.027 (flux systematics).
Including the data with kilometer baseline, the new world average is only about 1.4$\sigma$ lower than 
the unity, weakening the significance of the reactor antineutrino anomaly. 
The upcoming results from Daya Bay, RENO, and the Double Chooz will provide further 
information about this issue.

\end{abstract}

\maketitle
\thispagestyle{plain}

\section{Introduction}

The term ``reactor anomaly" was coined by Mention {\it et al.}~\cite{anom} who noted that the
average of the experimentally determined reactor antineutrino flux at
reactor-detector distances $<$ 100 m accounts for only 0.943 $\pm$ 0.023 of the reevaluated
theoretical expectation of Ref.~\cite{nspec2}. 
In addition to the 19 experimental results obtained with detectors distant less than 100 m from 
the reactor source, we are able now to include in the analysis also the results of the 
Chooz~\cite{Chooz1,Chooz2} and Palo Verde~\cite{PaloVerde} as well as of the 
Double Chooz~\cite{DChooz,DChooz1,DChooz2} 
experiments, where the detectors were further away from the reactor complex.  
In these cases, the corresponding experimental results need be corrected for the flux loss associated 
with the known value of the mixing angle $\theta_{13}$, which was determined in a model 
independent way by comparing the count rates in two essentially identical,
 \textcolor{black}{but separated in distance}, detectors. 
We use the value $\sin^2 2\theta_{13} = 0.089 \pm 0.011$ obtained in the Daya Bay 
experiment~\cite{dayabay,dayabay_long,dayabay_cpc}, and confirmed by the RENO experiment~\cite{reno}. 
The corresponding correction is easy to apply using the formula for the survival probability
\begin{widetext}
\begin{equation}
P_{sur} = 1- \sin^2 2\theta_{13} (\cos^2\theta_{12}\sin^2 \Delta_{31}+\sin^2\theta_{12}\sin^2{\Delta_{32}})
-\cos^4\theta_{13}\sin^2 2\theta_{12} \sin^2 \Delta_{21},
\end{equation}
\end{widetext}
with $\Delta_{ij} \equiv |\Delta_{ij}|= 1.27 |\Delta m^2_{ij}| \frac{L(m)}{E(MeV)}$. 
Values of mixing angles and mass-squared differences used in the simulation are 
taken from Ref.~\cite{PDG} assuming normal mass hierarchy ($\Delta m^2_{31} = \Delta m^2_{32} + \Delta m^2_{21}$):
\begin{eqnarray}
\sin^2 2\theta_{12} &=& 0.857 \pm 0.024 \nonumber\\
\Delta m^2_{21} &=& (7.50 \pm 0.20)\times 10^{-5} eV^2 \nonumber\\
\Delta m^2_{32} &=& (2.32 \pm 0.12)\times 10^{-3} eV^2.
\end{eqnarray}

\section{Analysis Description}

\begin{figure}[htb]
\centering
\includegraphics[width=90mm]{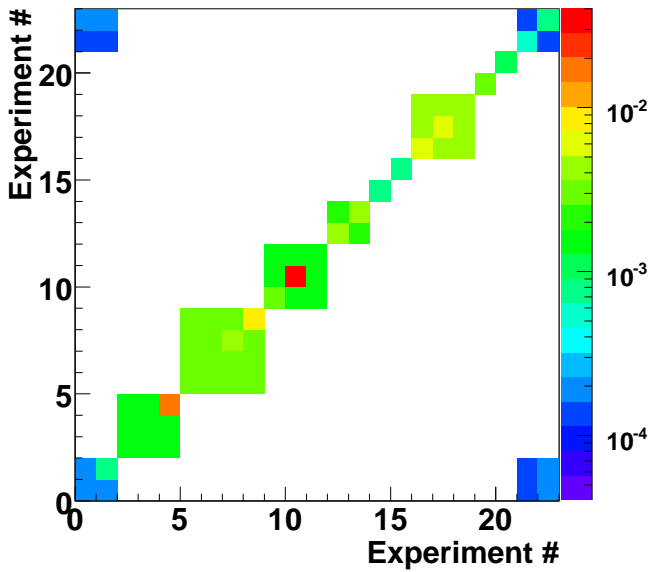}
\caption{(color online)
The covariance matrix of the reduced experimental uncertainties. The number of experiments can be 
found in Table.~\ref{table:results}. The off-diagonal terms show the correlation among different 
experiments.}
\label{fig:cov}
\end{figure}

In the following, we will explain in detail the inputs to our analysis as well as
the chi-square method. For the 19 experimental results obtained with detectors distant 
less than 100 m from the reactor source, the measured fluxes normalized to the prediction of 
Ref.~\cite{nspec2} (noted as ``ratio'') are taken from Ref.~\cite{anom} and tabulated in 
Table.~\ref{table:results}, together with the detector technology, fission fractions, 
distance to the reactor core, and year of publication. \textcolor{black}{The average survival 
probability $P_{sur}$ for each experiment is calculated by integrating over 
the neutrino antineutrino spectrum~\cite{nspec2} convoluted with the inverse beta decay cross
section with $\sin^22\theta_{13}=0.089$. }
 There are two uncertainties 
listed in Table~\ref{table:results}. The $\sigma_{err}$ represents the total uncertainty
on the ratio. The $\sigma_{corr}$ represents the part of uncertainty correlated among 
different experiments. In particular, there 
is a 2.7$\%$ uncertainty in $\sigma_{corr}$ coming from the uncertainty in the predicted
reactor flux. (\textcolor{black}{In first versions of Ref.~\cite{anom}, the reactor
flux uncertainty was assumed to be 2.7\%, which was replaced by 2\% in the final version. 
In this analysis, we chose the more conservative number (2.7\%).}) Since it will not affect the relative differences among different measurements,
we define the reduced uncertainties:
\begin{eqnarray}\label{eq:reduced}
\sigma^{reduced}_{err} = \sqrt{\sigma^2_{err} - 2.7^2} \nonumber \\
\sigma^{reduced}_{corr} = \sqrt{\sigma^2_{corr} - 2.7^2}
\end{eqnarray}
after removing the 2.7\% overall normalization uncertainty. 
\textcolor{black}{We change the $\sigma_{corr}$ between SRP-I and SRP-II~\cite{SRP}
from 3.7\% (originally quoted in 
Ref.~\cite{anom}) to 2.7\% (the reactor flux uncertainty only). 
The original 3.7\% assumes almost full correlation between SRP-I and SRP-II, which can not explain the 
apparent differences between the two ratios (0.952 vs. 1.018), indicating less correlation between 
the two experiments. For the same reason, we 
reduce the $\sigma_{corr}$  between 
the ROVNO88-1I and ROVNO88-2I from 6.9\% to 5.7\%, which 
is the final reported uncertainty for ROVNO88 
experiments~\cite{Rovno88}.}

There are three reactor cores in the Palo Verde experiment~\cite{PaloVerde}. The distances 
between detector and each reactor core are 750 m, 890 m, and 890 m. In calculating the average
survival probability $P_{sur}$, we assume that all three reactor cores have equal power.
The result is compared with $P^{750 m}_{sur}$ assuming full power in only the 750 m reactor 
and $P^{890 m}_{sur}$ assuming full power in the 890 m reactors. The differences are quoted as 
an additional uncertainty, which is  \textcolor{black}{only} about 5\% of the total 
reduced experimental uncertainty. 

There are two reactor cores in Chooz experiment~\cite{Chooz1,Chooz2}. The distances between 
the detector and each reactor core are 998 m and 1115 m. Similar procedure is applied to calculate
the uncertainty for the equal power assumption. The resulting uncertainty is 
about 6.2\% of the total reduced experimental uncertainty. The fission fractions are
assumed to be the same as those from Double Chooz~\cite{DChooz1}. We also calculated the 
average $P_{sur}$ by varying these fission fractions. The differences are negligible. 

The Double Chooz experiment is conducted at the same location as Chooz. 
With a single detector, the recent rate-only analyses of the data from delayed neutron 
capture on Gadolinium (n-Gd) and delayed neutron capture on hydrogen (n-H) reported the 
value of $\sin^22\theta_{13}=0.170 \pm 0.052$~\cite{DChooz1} 
and $\sin^22\theta_{13}=0.044 \pm 0.060$~\cite{DChooz2}~\footnote{To be consistent with 
other experiments, we choose the rate-only $\sin^22\theta_{13}$ results, which represent simple measures
of the disappearance in the total number of events.}, by anchoring to the short-baseline 
Bugey-4 results~\cite{Bugey4}, respectively. Although the measured flux normalized to the prediction of 
Ref.~\cite{nspec2} has not been reported, we can deduce such ratios using the reported
fission fractions~\cite{DChooz1}, the reported values of $\sin^22\theta_{13}$~\cite{DChooz1,DChooz2},
and the Bugey-4 results. The $\sigma^{reduced}_{err}$ is dominated by the uncertainties of 
reported $\sin^22\theta_{13}$, with additional uncertainties coming from the equal power
assumption. The $\sigma^{reduced}_{corr}$ are calculated from the reduced experimental uncertainty
$\sigma^{reduced}_{err}$ from Bugey-4. Furthermore, there are additional correlated uncertainties
between the n-H and n-Gd measurements due to the equal power assumption. 
The final covariance matrix $W$ using reduced uncertainties is shown in Fig.~\ref{fig:cov}.

The $\chi^2$ function used in this analysis is constructed as follows:
\begin{widetext}
\begin{equation}\label{eq:chi2}
\chi^2 (r,\sin^22\theta_{13}) = (r \cdot \vec{P}_{sur}(\sin^22\theta_{13})-\vec{R})^T W^{-1} (r \cdot\vec{P}_{sur}(\sin^22\theta_{13})-\vec{R}) 
+ \frac{(\sin^22\theta_{13}-0.089)^2}{0.011^2}.
\end{equation}
\end{widetext}
Here, $W^{-1}$ is the inverted covariance matrix. The vector $\vec{R}$ contains the reported ratios from 
all 23 experiments (tabulated in Table.~\ref{table:results}). \textcolor{black}{The absolute normalization ratio $r$ 
is treated as a free parameter.} 
The vector $\vec{P}_{sur}$ contains the predicted average survival probabilities given a value of $\sin^22\theta_{13}$.
The values of $\vec{P}_{sur}$ using $\sin^22\theta_{13}=0.089$ are tabulated in Table.~\ref{table:results}.
The last term in Eq.~\eqref{eq:chi2} represents the constrain on $\sin^22\theta_{13}$ from the latest Daya Bay 
results~\cite{dayabay_cpc}.

\section{Results}

\begin{figure}[htb]
\centering
\includegraphics[width=90mm]{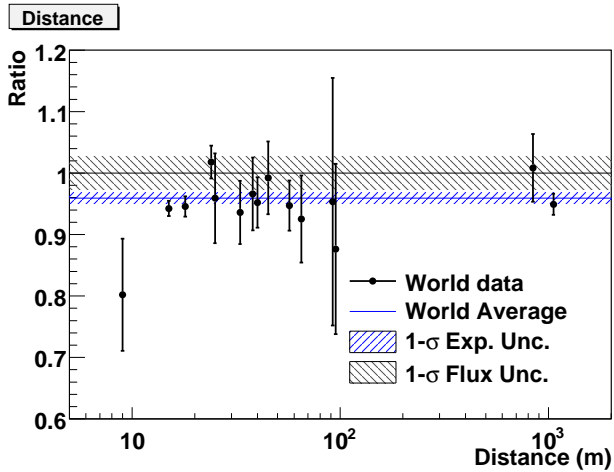}
\caption{ (color online) The reactor $\bar{\nu}_e$ capture rate as a function of the
distance from the reactor, normalized to the theoretical flux of Ref. \cite{nspec2}.
The horizontal bar represents the global average and its $1\sigma$ error bar. The
2.7\% reactor flux uncertainty is shown as a band around unity. We combine results at 
same baseline \textcolor{black}{(e.g. Chooz, Double Chooz n-H and n-Gd results)} together for clarity.
}
\label{fig:dist}
\end{figure}

In the following three figures, we show the results of all 23 measurements 
and the deduced ratios after minimizing the $\chi^2$ defined in Eq.~\eqref{eq:chi2}.
The global average is determined to be 0.959$\pm 0.009$. In Fig. \ref{fig:dist}, the 
results are shown in analogous way as in Ref. \cite{anom}, i.e. as a function of the 
distance from the corresponding reactor core. We combine results at same baseline 
together for clarity. The corresponding $\chi^2$/Dof = 23.8/22. We stress that our error 
bars do not include the reactor flux uncertainty (2.7\%), hence they appear smaller 
than those in~\cite{anom}. 

\begin{figure}[htb]
\centering
\includegraphics[width=90mm]{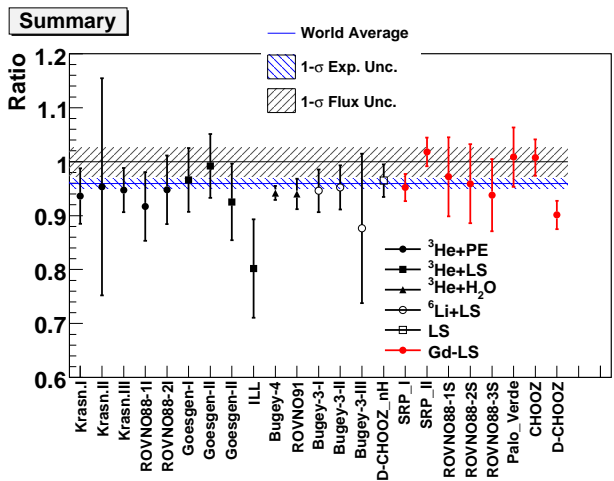}
\caption{ (color online)  The same 23 experimental results as in previous figures are plotted by the detection 
method employed. See captions of Fig.~\ref{fig:dist}
for details.
}
\label{fig:twogroups}
\end{figure}

\textcolor{black}{The new global average is somewhat larger than the 0.943 value 
of Ref.\cite{anom} quoted earlier, weakening the significance of the reactor 
antineutrino anomaly. There are two reasons for this difference. First, 
we include more recent and more distant experiments, of which Palo Verde and Chooz 
have larger rates. Second, we change the correlated uncertainty between SRP-I and SRP-II
from 3.7\% to 2.7\% (the reactor flux uncertainty only), since the original 3.7\% 
assumes almost full correlation between SRP-I and SRP-II, which can not explain the 
apparent differences between the two ratios (0.952 vs. 1.018). 
With fixed total experimental uncertainties, this change effectively increases the 
significance of SRP experiments and leads to about 1\% larger world average. }

\textcolor{black}{
In addition, our results are larger than those reported in Ref.~\cite{Evslin} and 
Ref.~\cite{thomas_new}, which also include the kilometer 
experiments with known $\theta_{13}$. The result reported in Ref.~\cite{Evslin}
included the Gallium neutrino data~\cite{GALLEX1,GALLEX2,SAGE1,SAGE11,SAGE2,SAGE3,Gallium3}, 
which was not included in our reactor antineutrino analysis. They also did an analysis
by including a RENO preliminary result 
from the absolute flux analysis. However, 
such analysis of the RENO experiment has not been, 
to our knowledge, released and is not finished as yet~\cite{soobong}.
In Ref.~\cite{thomas_new}, the measured experimental fluxes are normalized to 
the predicted flux of Huber~\cite{nspec1} with a new neutron lifetime 881.5s 
(2011 update of PDG~\cite{PDG}). \textcolor{black}{The change in reactor flux model and 
the neutron lifetime leads to in average 1.6\% lower ratios
than what we used in this work (tabulated in Table.~\ref{table:results}).} 
The rest of differences come from the treatment in the correlated uncertainty of SRP 
experiments and the uncertainty of the reactor flux prediction (2\% used in 
Ref.~\cite{thomas_new} vs. 2.7\% used in this work).}

\textcolor{black}{
One of the main purposes of this work is to illustrate the impact of kilometer experiments 
to the results of Mention {\it et al.}~\cite{anom}. Therefore, we have adapted the same neutron 
life time (885.7s) as in Mention {\it et al.}~\cite{anom}. The current recommended neutron lifetime 
from the 2012 Particle Data Group~\cite{PDG} is 880.1s. Using the latest neutron life time would
lead to about 0.63\% reduction in the average ratio.  
}

\begin{figure}[htb]
\centering
\includegraphics[width=90mm]{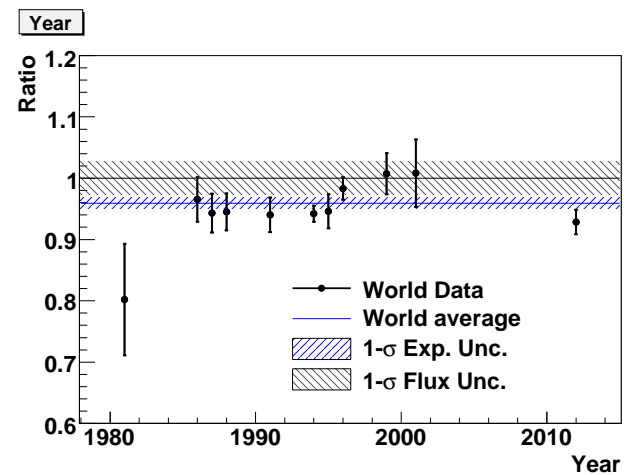}
\caption{ (color online)  The reactor $\bar{\nu}_e$ capture rate as a function of the
year when the measurement was published, normalized to the theoretical flux of Ref. \cite{nspec2}. We combine results at 
same year \textcolor{black}{(e.g. Double Chooz n-H and n-Gd results)} together for clarity. See captions of Fig.~\ref{fig:dist}
for details.
}
\label{fig:year}
\end{figure}

\begin{figure}[htb]
\centering
\includegraphics[width=90mm]{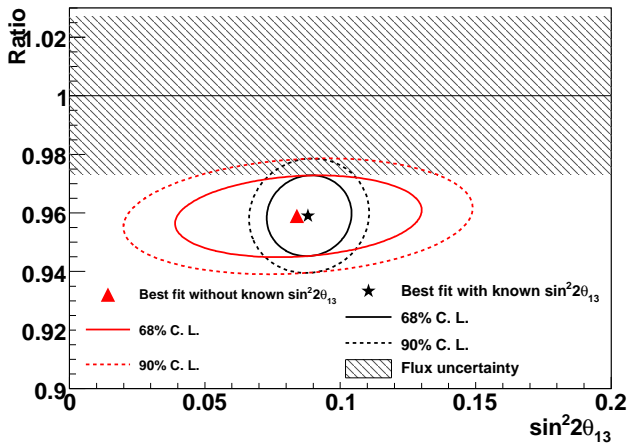}
\caption{ (color online) \textcolor{black}{Best fits are shown with and without $\sin^22\theta_{13}$
constrains from Daya Bay. The corresponding 68\% ($\Delta \chi^2 < 2.3$) and 
90\% ($\Delta \chi^2<4.6$) confidence intervals are shown as well. }
}
\label{fig:cont}
\end{figure}

Fig.~\ref{fig:twogroups} shows results from all 23 experiments again arranged by the detector technology. 
Five different technologies were used in the 23 experiments to record the 
 $\bar{\nu}_e$ capture on protons. In the ILL~\cite{ILL} and Goesgen experiments \cite{Goes} the 
liquid scintillator targets cells were interspaced with the $^3$He neutron counters.
In the Bugey-3 experiments \cite{Bugey3_1,Bugey3_2} the liquid scintillator was loaded with $^6$Li 
to detect the neutron captures. In the Bugey-4~\cite{Bugey4} and Rovno91 \cite{Rovno91} only the 
neutron captures were detected. The detector in these two experiments consisted of a water target 
with embedded $^3$He detectors. In the Krasnoyarsk \cite{Krasn} and Rovno88 \cite{Rovno88} experiments 
again only the total capture rates were measured. The detectors consisted of the polyethylene neutron 
moderator with $^3$He neutron counters embedded in them. Finally, the Savannah River experiments~\cite{SRP}, 
the Rovno88 \cite{Rovno88} and the Chooz, Palo Verde and Double Chooz experiments~\cite{DChooz1}~\footnote{
In Double Chooz n-H analysis~\cite{DChooz2}, the neutrino events are largely generated in the liquid 
scintillator region without the Gd loading.} use the Gd loaded liquid scintillators. The Daya Bay and RENO 
experiments are employing that technology as well. The results from the Gd loaded 
scintillator experiments are, with the exception of Double Chooz,  in general higher 
than the new world average. 

\textcolor{black}{We should also note that the experiments with Gd loaded scintillator 
were in general carried out at a later time.
In Fig. \ref{fig:year} we show the same data, but now arranged as a function of the year 
when the corresponding measurements were published. One can see that the more recent 
experiments, except the Double Chooz, appear to have higher rates than 
the earlier ones. If such tendency is true, it could either be due to difference 
in detector technologies or improvement in data acquisition or analysis methods. However, 
the $\chi^2$ of the global fit, $\chi^2$/Dof = 23.8/22, signifies that the 23 reactor flux 
determinations are mutually consistent. Therefore, our observation might be
simply due to statistical fluctuations.}

In addition, we also use these 23 experiments to extract $\sin^22\theta_{13}$ 
by minimizing the $\chi^2$ in Eq.~\eqref{eq:chi2} without the penalty term 
($\frac{(\sin^22\theta_{13}-0.089)^2}{0.011^2}$). 
The best-fit $\sin^22\theta_{13}$ is determined to be 0.084 $\pm$ 0.030 
(one dimension: $\Delta \chi^2 \equiv \chi^2 - \chi^2_{minimum}<1$). 
\textcolor{black}{The non-zero $\sin^22\theta_{13}$ is dominated by the latest
Double Chooz results, and is also consistent with the latest Daya Bay 
results~\cite{dayabay_cpc}. } Fig.~\ref{fig:cont} shows the 
best-fit $\sin^22\theta_{13}$ and ratio $r$ with and without the 
Daya Bay constrains. The corresponding 68\% (two dimensions: $\Delta \chi^2 < 2.3$) 
and 90\% (two dimensions: $\Delta \chi^2<4.6$) confidence intervals are shown as well. 
\textcolor{black}{The correlation between the ratio and $\sin^22\theta_{13}$ is rather weak
due to the strong correlation between results from Double Chooz, which dominate
the non-zero $\sin^22\theta_{13}$ extraction from these 23 experiments, and result from 
Bugey-4, which is the most precise short baseline measurement. 
Therefore, it is not surprising that the precise $\sin^22\theta_{13}$ value from 
Daya Bay, which is very close to the best fit $\sin^22\theta_{13}$ value from these 23 
experiments, does not improve the knowledge on the ratio.}

\section{Summary and Discussion}

\textcolor{black}{With the known $\theta_{13}$, we include results from Palo Verde, Chooz, 
and Double Chooz, and re-evaluate the reactor antineutrino anomaly. The new world average
is determined to be 0.959 $\pm$ 0.009 (experiment uncertainty) 
$\pm$ 0.027 (flux systematics), when normalized to the predicted flux of 
Mueller {\it et al.}~\cite{nspec2}.} The new world average is about 1.4$\sigma$ lower 
than unity, weakening the significance of the reactor antineutrino anomaly. 
We also show that the analysis of all 23 experiments yields $\sin^22\theta_{13}=0.084\pm0.030$ 
in agreement with its independently obtained value from the Daya Bay and RENO experiments. 

\textcolor{black}{The forthcoming absolute flux analysis of the Daya Bay, RENO, and Double Chooz experiments 
will clearly shed more light on this issue. The final answer to the question of agreement between 
the predicted reactor $\bar{\nu}_e$ flux and the measured rate clearly depends on the results 
of these experiments as well as on the careful analysis of the predicted flux and its uncertainties. }

\textcolor{black}{The most popular explanation of the anomaly, if it turns out that it is statistically 
significant, is the existence of additional sterile neutrinos. The consistency of the rate of the short 
baseline neutrino experiments, demonstrated here and also e.g. in Ref.~\cite{anom,thomas_new} suggests 
that such additional neutrinos must have large enough masses that the corresponding oscillation length
is at most few meters. 
Experiments sensitive to such short oscillation length would be able to convincingly prove the existence of 
the sterile neutrino and allow determination of their masses and mixing angles. 
}


\begin{table*}[ht!]
\begin{center}
\begin{tabular}{|c|c|c|c|c|c|c|c|c|c|c|c|c|}
\hline
\hline
\# & result     & Det. type         & $^{235}$U & $^{239}$Pu & $^{238}$U & $^{241}$Pu & ratio & $\sigma_{err}$(\%) & $\sigma_{corr}$ (\%)&  L(m) & $P_{sur}$   & Year \\\hline
1 & Bugey-4    & $^3$He + H$_2$O   & 0.538     & 0.328      & 0.078     & 0.056      & ~0.942~ & 3.0     & 3.0      & 15     &  ~0.999987~   &  ~1994\\
2 & ROVNO91    & $^3$He + H$_2$O   & 0.614     & 0.274      & 0.074     & 0.038      & ~0.940~ & 3.9     & 3.0      & 18     &  ~0.999981~   &  ~1991\\
22&Double Chooz & Gd-LS             & 0.496     & 0.351      & 0.087     & 0.066      & ~0.860~ & 3.7     & 3.0      & 998-1115& ~0.954~   &  ~2012\\
23&Double Chooz & LS (n-H)          & 0.496     & 0.351      & 0.087     & 0.066      & ~0.920~ & 4.0     & 3.0      & 998-1115& ~0.954~   &  ~2012\\\hline
3 & Bugey-3-I  & $^6$Li - LS       & 0.538     & 0.328      & 0.078     & 0.056      & ~0.946~ & 4.8     & 4.8      & 15     &  ~0.999987~   &  ~1995\\
4 & Bugey-3-II & $^6$Li - LS       & 0.538     & 0.328      & 0.078     & 0.056      & ~0.952~ & 4.9     & 4.8      & 40     &  ~0.999907~   &  ~1995\\
5 & Bugey-3-III& $^6$Li - LS       & 0.538     & 0.328      & 0.078     & 0.056      & ~0.876~ & 14.1    & 4.8      & 95     &  ~0.999479~   &  ~1995\\\hline
6 & Goesgen-I  & $^3$He + LS       & 0.620     & 0.274      & 0.074     & 0.042      & ~0.966~ & 6.5     & 6.0      & 38     &  ~0.999916~   &  ~1986\\
7 & Goesgen-II & $^3$He + LS       & 0.584     & 0.298      & 0.068     & 0.050      & ~0.992~ & 6.5     & 6.0      & 45     &  ~0.999883~   &  ~1986\\
8 & Goesgen-III& $^3$He + LS       & 0.543     & 0.329      & 0.070     & 0.058      & ~0.925~ & 7.6     & 6.0      & 65     &  ~0.999756~   &  ~1986\\
9 & ILL        & $^3$He + LS       & $\approx$1& -          & -         & -          & ~0.802~ & 9.5     & 6.0      & 9      &  ~0.999995~   &  ~1981\\\hline
10& Krasn. I   & $^3$He + PE       & $\approx$1& -          & -         & -          & ~0.936~ & 5.8     & 4.9      & 33     &  ~0.999937~   &  ~1987\\
11& Krasn. II  & $^3$He + PE       & $\approx$1& -          & -         & -          & ~0.953~ & 20.3    & 4.9      & 92     &  ~0.999511~   &  ~1987\\ 
12& Krasn. III & $^3$He + PE       & $\approx$1& -          & -         & -          & ~0.947~ & 4.9     & 4.9      & 57     &  ~0.999812~   &  ~1987\\\hline
13& SRP-I      & Gd-LS             & $\approx$1& -          & -         & -          & ~0.952~ & 3.7     & 2.7 & 18     &  ~0.999981~   &  ~1996\\
14& SRP-II     & Gd-LS             & $\approx$1& -          & -         & -          & ~1.018~ & 3.8     & 2.7 & 24     &  ~0.999967~   &  ~1996\\\hline
15& ROVNO88-1I & $^3$He + PE       & 0.607     & 0.277      & 0.074     & 0.042      & ~0.917~ & 6.9     & 5.7 & 18     &  ~0.999981~   &  ~1988\\
16& ROVNO88-2I & $^3$He + PE       & 0.603     & 0.276      & 0.076     & 0.045      & ~0.948~ & 6.9     & 5.7 & 18     &  ~0.999981~   &  ~1988\\\hline
17& ROVNO88-1S & Gd-LS             & 0.606     & 0.277      & 0.074     & 0.043      & ~0.972~ & 7.8     & 7.2      & 18     &  ~0.999981~   &  ~1988\\
18& ROVNO88-2S & Gd-LS             & 0.557     & 0.313      & 0.076     & 0.054      & ~0.959~ & 7.8     & 7.2      & 25     &  ~0.999964~   &  ~1988\\
19& ROVNO88-3S & Gd-LS             & 0.606     & 0.274      & 0.074     & 0.046      & ~0.938~ & 7.2     & 7.2      & 18     &  ~0.999981~   &  ~1988\\\hline
20& Palo Verde & Gd-LS             & 0.60      & 0.27       & 0.07      & 0.06       & ~0.975~ & 6.0     & 2.7      & 750-890&  ~0.967~   &  ~2001\\\hline
21& Chooz      & Gd-LS             & 0.496     & 0.351      & 0.087     & 0.066      & ~0.961~ & 4.2     & 2.7      & 998-1115& ~0.954~   &  ~1999\\\hline
\hline
\end{tabular}
\end{center}
\caption{\label{table:results} Tabulated results of all 23 experiments. Experiments are categorized into different groups 
with horizontal lines. Within each group, the $\sigma_{corr}$ represent the correlated uncertainties among different 
experiments. \textcolor{black}{This table is an extension of Table.~II of Ref.~\cite{anom}.
There are additional correlated uncertainties, since Double Chooz results were anchored to the Bugey-4. 
See the text for more explanations.}}
\end{table*}

\bigskip

{\large Acknowledgments}
We would like to thank R. D. McKeown and W. Wang for fruitful discussions. 
This work was supported in part by Caltech, the National Science
Foundation, and the Department of Energy  under contracts DE-AC02-98CH10886.

\bibliographystyle{unsrt}
\bibliography{abs_flux}{}

\end{document}